\documentclass[11pt]{article}
\usepackage[top=1in,bottom=1in,left=1in,right=1in]{geometry}

\usepackage[noEucal]{main}
\usepackage{aas_macros}
\usepackage{graphicx}

\def\[{\big[}
\def\]{\big]}
\def\la{\langle}
\def\ra{\rangle}

\def\ci{{\mathcal I}}

\def\dt{{\msd}}

\def\BX{{\mathbf X}}
\def\BY{{\mathbf Y}}

\def\Bi{{\mathbf i}}

\def\Bdt{{\mathbf{d}}}
\def\BOm{{\mathbf{\O}}}

\def\BTh{{\mathbf{\Theta}}}

\def\mfS{{\mathfrak S}}

\def\mfD{{\mathfrak D}}
\def\mfg{{\mathfrak g}}
\def\mfF{{\mathfrak F}}

\def\cc{{\text{c.c.}}}

\def\tA{{\bar \msA}}

\def\tF{{\bar \msF}}

\def\Nfield{{N}}

\begin{document}
\begin{titlepage}
\unitlength = 1mm

\hfill CALT-TH 2023-015

\vskip 3cm
\begin{center}

{\LARGE{\textsc{Asymptotic Structure of Higher Dimensional Yang-Mills Theory}}}

\vspace{0.8cm}
Temple He$^\ddagger$, Prahar Mitra$^{\Diamond}$

\vspace{1cm}

{\it  $^\ddagger$Walter Burke Institute for Theoretical Physics, California Institute of Technology, \\ Pasadena, CA 91125 USA}\\
{\it  $^\Diamond$Department of Applied Mathematics and Theoretical Physics, University of Cambridge,\\ Wilberforce Road, Cambridge CB3 0WA, UK}

\vspace{0.8cm}

\begin{abstract}

Using the covariant phase space formalism, we construct the phase space for non-Abelian gauge theories in $(d+2)$-dimensional Minkowski spacetime for any $d \geq 2$, including the edge modes that symplectically pair to the low energy degrees of freedom of the gauge field. Despite the fact that the symplectic form in odd and even-dimensional spacetimes appear ostensibly different, we demonstrate that both cases can be treated in a unified manner by utilizing the shadow transform. Upon quantization, we recover the algebra of the vacuum sector of the Hilbert space and derive a Ward identity that implies the leading soft gluon theorem in $(d+2)$-dimensional spacetime.
\end{abstract}

\vspace{1.0cm}
\end{center}
\end{titlepage}
\pagestyle{empty}
\pagestyle{plain}
\pagenumbering{arabic}

\tableofcontents

\normalem

\section{Introduction}

In recent years, the discovery of the equivalence between asymptotic symmetries, soft theorems, and the memory effect has led to a resurgence in the study of the infrared sector of quantum field theories (QFTs) (for a review, see \cite{Strominger:2017zoo}, and references therein). This triangle of equivalence, dubbed the infrared triangle, was first shown in the context of four-dimensional gravity in asymptotically flat spacetimes \cite{Strominger:2013jfa,He:2014laa,Strominger:2014pwa}, where it was established that the leading soft graviton theorem \cite{Weinberg:1965aa} is the Ward identity for BMS supertranslations \cite{Bondi:1962px,Sachs:1962wk}, which in turn is associated to a gravitational memory effect \cite{Braginsky:1986ia,Wiseman:1991ss,Christodoulou:1991cr,Thorne:1992sdb}.
This was later extended to four-dimensional gauge theories \cite{He:2014cra,He:2015zea, Pasterski:2015zua, Campiglia:2015qka, Dumitrescu:2015fej, Pate:2017vwa, Ball:2018prg}, as well as higher even-dimensional gravity and gauge theories \cite{Kapec:2014zla, Kapec:2015vwa, Pate:2017fgt}. Furthermore, beyond the leading soft theorems, there has also been extensive research connecting the subleading (and sub-subleading) soft theorems to asymptotic symmetries and memory effects, e.g. see \cite{Campiglia:2014xja,Kapec:2014opa,Lysov:2014csa,Pasterski:2015tva, Campiglia:2015kxa, Campiglia:2016hvg, Campiglia:2016jdj}.

Although soft theorems exist in all spacetime dimensions, it was only recently that the connection between soft theorems and asymptotic symmetries was established in odd spacetime dimensions. The main technical challenge was due to the qualitatively different properties of massless wave propagation in odd and even-dimensional spacetimes. This difference is often referred to as the failure of Huygens' principle in odd-dimensional spacetimes. For the case of massless gauge theories, this difficulty was surmounted in \cite{He:2019jjk,He:2019pll,He:2019ywq}, where it was demonstrated that the charge generating large gauge transforms (LGTs) in odd-dimensional spacetime has a somewhat different form from those in even-dimensional spacetime.

Meanwhile, in order to study the vacuum sector of the Hilbert space in gauge theories, an analysis was carried out in \cite{He:2020ifr} using the covariant phase space formalism to construct the symplectic form of four-dimensional non-Abelian gauge theories. This allowed for a derivation of the Dirac brackets pertaining to the soft modes, i.e. the soft gluon mode $N$ and its symplectic partner $C$, which is the gauge field edge mode. Upon canonical quantization, this led unambiguously to the algebra of soft operators in the vacuum Hilbert space, and the corresponding Ward identity was shown to give rise to the leading soft gluon theorem. In addition, it was identified that the charge generating LGTs are canonical transformations preserving the symplectic form.

The fact that the large gauge charge takes on different forms in odd and even spacetime dimensions suggests that it might be difficult to extend the analysis of \cite{He:2020ifr} in a uniform manner to both odd and even-dimensional non-Abelian gauge theories. However, inspired by the analysis of \cite{Kapec:2021eug}, where it was shown that the soft effective action of gravity and gauge theories in any dimension does not involve the soft operator, but rather its \emph{shadow transform}, we are led to wonder if it is perhaps more natural to write the symplectic form not in terms of the soft gluon operator $N$, but rather its shadow transform $\widetilde N$. We demonstrate in this paper that this is indeed the case. By writing the symplectic form in terms of the shadow transform of the soft operator, all differences between odd and even dimensions disappear, leading us to a uniform treatment of both odd and even-dimensional non-Abelian gauge theories via the covariant phase space formalism. This extends the analysis using covariant phase space formalism initiated in \cite{He:2020ifr} to theories with dimensions greater than four, and also confirms the results of \cite{Kapec:2022hih} pertaining to non-Abelian gauge theories.

Because this paper is a direct extension of the analysis performed in \cite{He:2020ifr}, we will oftentimes neglect details and refer the reader to \cite{He:2020ifr} for a more in-depth analysis and treatment. In Section~\ref{sec:phase-space}, we use the covariant phase space formalism to construct the symplectic form for all $(d+2)$-dimensional theories for $d \geq 2$. In Section~\ref{sec:quantization}, we canonically quantize the phase space and obtain the quantum commutators by inverting the symplectic form, and then construct the Hilbert space, including both the vacuum sector as well as the radiative Fock space. Finally, in Section~\ref{sec:wardid}, we write down the Ward identity associated with LGTs and use it to derive the leading soft gluon theorem.

\section{Asymptotic Phase Space}\label{sec:phase-space}

In this paper, we use the covariant phase space formalism to construct the symplectic form for non-Abelian gauge theories in $d+2$ dimensions. The notations and conventions employed in this paper are those used in \cite{He:2020ifr}. For completeness, we review some of them in Appendix \ref{app:conventions} and introduce others as needed in the main text, but for more details, we refer the reader to \cite{He:2020ifr}.

\subsection{Covariant Phase Space}

Consider a non-Abelian gauge theory in a $(d+2)$-dimensional spacetime $(\CM,g)$, where $\CM$ is the Lorentzian spacetime manifold and $g$ the metric, with compact semi-simple gauge group $\CG$ and corresponding Lie algebra $\mfg$. On $(\CM,g)$, we set up a generalized coordinate system $\{x^\mu\}$, and on $\mfg$, we install a basis of generators $T^I$, where $I \in \{1,\ldots,|\mfg|\}$, that satisfy $[ T^I , T^J ] = f^{IJK} T^K$. The dynamics of a non-Abelian gauge field $\msA = A_\mu^I(x)\, \msd x^\mu \otimes T^I$ in $d+2$ dimensions is described minimally by a Lagrangian form
\begin{equation}
\begin{split}\label{YM-Lagrangian}
\msL = \frac{1}{2g^2} \Tr [ \msF \wedge \star \msF ] , \qquad \msF = \msd \msA + \msA \wedge \msA , 
\end{split}
\end{equation}
where $\Tr$ is the trace in the adjoint representation normalized so that $\Tr[ T^I T^J] = -\d^{IJ}$. More generally, we could also add higher derivative terms, as well as terms involving other matter fields, to the Lagrangian \eqref{YM-Lagrangian}. This generalization was studied in great detail in \cite{He:2020ifr}, and as those extra terms do not play a role in the structure of the phase space on $\ci^\pm$, we will safely ignore them for the rest of our analysis.

The Lagrangian \eqref{YM-Lagrangian} is a $(d+2)$-form in spacetime and a 0-form (function) in field configuration space $\mfF$.\footnote{This is the space of \emph{allowed} field configurations of the gauge field $\msA$, which is defined by choosing appropriate boundary conditions.} A generic vector field ${\bf X}$ on $\mfF$ is given by
\begin{equation}
\begin{split}
\BX \equiv \int_\CM \dt^{d+2} x\, \sqrt{-g} \, \d A_\mu^I(x) \frac{\d}{\d A_\mu^I(x)}  ,
\end{split}
\end{equation}
and the action of such a vector on $\msL$ is
\begin{align}\label{varL}
\begin{split}
\BX ( \msL ) &= \frac{1}{g^2} \Tr \left[ \BX(\msA) \wedge \left( \msd \star \msF + \msA \wedge \star \msF - (-1)^d \star \msF \wedge \msA \right) \right] + \msd \bigg( \frac{1}{g^2} \Tr[\BX(\msA) \wedge \star \msF] \bigg).
\end{split}
\end{align}
The first term is used to define the \emph{solution space} $\mfS$, which is the subspace of $\mfF$
\begin{equation}
\begin{split}
\mfS \equiv \big\{ \msA \in \mfF ~|~  \msd \star \msF + \msA \wedge \star \msF - (-1)^d \star \msF \wedge \msA  = 0 \big\} . 
\end{split}
\end{equation}
In the rest of this paper, we will work exclusively on $\mfS$, which is known as going \emph{on-shell}.

In the covariant phase space formalism \cite{Crnkovic:1986ex, Lee:1990nz, Wald:1999wa, Iyer:1994ys, Harlow:2019yfa, He:2020ifr}, the pre-symplectic potential is constructed by integrating the second term in \eqref{varL} over a Cauchy slice $\S$, i.e.
\begin{align}
\label{symp_pot0}
\begin{split}
\wt{\BTh}_\S(\BX) = \frac{1}{g^2} \int_\S \Tr[\BX(\msA) \wedge \star \msF ].
\end{split}
\end{align}
The pre-symplectic form is then constructed by taking an exterior derivative of this on $\mfS$, and is given by
\begin{equation}
\begin{split}
\label{symp_form0}
\wt{\BOm}_\S (\BX,\BY) = - \frac{1}{g^2} \int_\S \Tr[ \BX(\msA) \wedge \star \BY (\msF )   - \BY (\msA) \wedge \star \BX (\msF )   ] . 
\end{split}
\end{equation}
Without any specific choice of boundary conditions for the gauge field on $\p\S$, the pre-symplectic form generically depends on the choice of $\S$. This can be fixed (as shown in \cite{He:2020ifr}) by restricting to a solution space satisfying 
\begin{equation}
\begin{split}\label{C-def}
    \msF \big|_{\p\S} = 0 \quad \implies \quad \msA \big|_{\p\S} = C \msd C^{-1} , \qquad C \in \CG. 
\end{split}
\end{equation}
On this solution space, it is convenient to write the gauge field as\footnote{Note that \eqref{C-def} defines $C$ only on $\p\S$. To decompose the gauge field as in \eqref{gauge-field-decomp}, we need to smoothly extend $C$ into $\S$. This extension is not unique, but as we will see from \eqref{symp_form}, the bulk extension of $C$ does not enter the symplectic form, so all extensions are physically equivalent.}
\begin{align}
\label{gauge-field-decomp}
\msA = C \tA C^{-1} + C \msd C^{-1} , \qquad \tA|_{\p\S} = 0.
\end{align}
Substituting \eqref{gauge-field-decomp} into \eqref{symp_pot0}, we find that the symplectic form takes the form
\begin{align}
\label{symp_form1}
\begin{split}
\wt{\BOm}(\BX,\BY) &= - \frac{1}{g^2} \int_\S \Tr[ \BX(\tA) \wedge \star \BY(\tF)  - \BY(\tA) \wedge \star \BX(\tF)  ] \\
&\qquad \qquad \qquad + \frac{1}{g^2} \oint_{\p\S} \Tr [ \BX(C) \star \BY(\tF C^{-1} ) - \BY(C) \star \BX(\tF C^{-1} )  ]  , 
\end{split}
\end{align}
where $\tF = \dt \tA + \tA \wedge \tA$. Note that we have now dropped the $\S$ subscript on ${\wt \BOm}$.

Given the pre-symplectic form \eqref{symp_form1}, the phase space of the theory is given by $\G = \mfS / \ker \wt{\BOm}$, and the corresponding symplectic form is defined by $\BOm = \wt{\BOm}|_\G$. One class of elements of the kernel is given by the vector field that generates infinitesimal gauge transformations\footnote{Finite gauge transformations act on the gauge field as $\msA \to g \msA g^{-1} + g \msd g^{-1}$, or equivalently $\tA \to \tA$, $C \to g C$.}
\begin{equation}
\begin{split}
    \BX_\ve = - \int_\CM \dt^{d+2} x \,\sqrt{-g} \,( D_\mu \ve)^I \frac{\d}{\d A_\mu^I} , \qquad D_\mu \ve \equiv \p_\mu \ve + [ A_\mu , \ve ] . 
\end{split}
\end{equation}
It can be verified via direct computation that the action of this vector on $\tA$ and $C$ is given by
\begin{equation}
\begin{split}
\BX_\ve(\tA) = 0 , \qquad \BX_\ve(C) = \ve C . 
\end{split}
\end{equation}
In particular, we see that $\tA$ is gauge invariant. Using this, we find
\begin{equation}
\begin{split}\label{Om-null}
\wt{\BOm}(\BY,\BX_\ve) &= - \frac{1}{g^2} \oint_{\p\S} \Tr [ \ve \star \BY( \msF )  ]  .
\end{split}
\end{equation}
We see from this that if $\ve |_{\p\S} = 0$, then $\BX_\ve$ lives in the kernel of ${\wt \BOm}$. Therefore, these are the \emph{small gauge transformations}, and the phase space $\G$ is constructed by identifying all the transformations generated by this vector. In practice, this is achieved by imposing a gauge fixing condition $f[A_\mu]=0$ that removes these redundancies.

On the other hand, we see from \eqref{Om-null} that if $\ve|_{\p\S} \neq 0$ and $\ve$ is field-independent, i.e. $\BY(\ve) = 0$, then we can rewrite \eqref{Om-null} as
\begin{align}
    \wt{\BOm}(\BY,\BX_\ve) = \BY \bigg( - \frac{1}{g^2} \oint_{\p\S} \Tr[\ve \star \msF ] \bigg),
\end{align}
which means $\BX_\ve$ is a Hamiltonian vector field that generates a canonical transformation. These are known as \emph{large gauge transformations} (LGTs), and the corresponding Hamiltonian charge is
\begin{equation}
\begin{split}
\label{large_gauge_charge}
Q_\ve =  -  \frac{1}{g^2} \oint_{\p\S} \Tr [ \ve \star \msF ] . 
\end{split}
\end{equation}
Upon canonical quantization, the large gauge charge obeys the quantum commutation relation\footnote{The quantum commutator is related to the Poisson/Dirac bracket by a factor of $i$, such that $[\cdot,\cdot] = i \{\cdot,\cdot\}$. \label{fn:dirac-comm}}
\begin{equation}
\begin{split}
\label{large_gauge_charge2}
    \left[ Q_\ve , Q_{\ve'} \right] = - i \BOm ( \BX_\ve , \BX_{\ve'} ) =  i Q_{[ \ve,  \ve' ]} . 
\end{split}
\end{equation}

\subsection{Symplectic Form on \texorpdfstring{$\ci^\pm$}{ipm}}

Thus far, our results have been rather general. We now focus on the case of interest. We take $\CM$ to be $(d\!+\!2)$-dimensional Minkowski spacetime and consider the symplectic form on two distinct Cauchy slices $\S^\pm = \ci^\pm \cup i^\pm$ with boundaries $\p\S^\pm = \ci^\pm_\mp$, where $\ci^+$ ($\ci^-$) is the future (past) null infinity, $i^+$ ($i^-$) the future (past) timelike infinity, and $\ci^+_-$ ($\ci^-_+$) the past (future) boundary of $\ci^+$ ($\ci^-$). To describe these surfaces, we will work in \emph{flat null coordinates} $(u,x^a,r)$, which are related to the usual Cartesian coordinates via
\begin{align}
\label{flat_null_coord}
X^\mu = \frac{r}{2} \bigg( 1 + x^2 + \frac{u}{r}, 2x^a, 1-x^2 - \frac{u}{r} \bigg) , \qquad x^2 = \d_{ab} x^a x^b,
\end{align}
where $\mu \in \{0,1,\cdots,d+1\}$ are the spacetime coordinates and $a,b \in \{1,\cdots,d\}$ are the transverse spatial coordinates.\footnote{We will throughout this paper use lowercase Greek indices to denote spacetime coordinates, and lowercase Latin indices to denote transverse spatial coordinates.} Transverse spatial indices are raised and lowered by the Kronecker delta $\d^{ab}$ and $\d_{ab}$, respectively. In flat null coordinates, Minkowski spacetime is then given by the metric
\begin{align}
    \dt s^2 = -\dt u \,\dt r + r^2 \delta_{ab} \,\dt x^a \dt x^b.    
\end{align}
The asymptotic null boundaries $\ci^\pm$ of Minkowski spacetime are located at $r \to \pm\infty$ with $(u,x^a)$ fixed. Their respective past and future boundaries $\ci^\pm_\mp$ are located by further taking the limit $u \to \mp\infty$. For more details, we refer the reader to Appendix A of \cite{He:2019jjk}.

With the identification $\S^\pm = \ci^\pm \cup i^\pm$, the boundary condition \eqref{C-def} is then the condition
\begin{align}\label{matching}
\msA|_{\ci^+_-} = \msA|_{\ci^-_+} = C\dt C^{-1},    
\end{align}
so the gauge field is flat and identified across spatial infinity $i^0$.\footnote{Note that the coordinate $x^a$ describes anti-podal points on $\ci^+$ and $\ci^-$, so \eqref{matching} is actually an \emph{anti-podal} matching condition.} This is the matching condition that has been imposed in previous works, e.g. \cite{Strominger:2014pwa, He:2014laa, He:2019jjk, He:2019pll}, and as argued in \cite{He:2020ifr} is sufficient for the symplectic form on $\ci^+$ to be the same as that on $\ci^-$.\footnote{Thus, we do not need to include $\pm$ superscripts on the field $C$.} This is equivalent to the requirement that there is no flux leaving the system through spatial infinity $i^0$.

To define the phase space with an invertible symplectic form, we need to fix the small gauge symmetry. Following \cite{He:2020ifr}, we do this by setting $A_u = 0$. The symplectic form \eqref{symp_form1} on $\S^\pm$ in flat null coordinates then works out to be
\begin{align}
\label{symp_form}
\begin{split}
\BOm(\BX,\BY) &=  \frac{1}{g^2} \int  \dt u \,\dt^d x \, \Tr\big[  \BX( {\hat A}^{\pm a} ) \BY( \p_u {\hat A}^\pm_a ) - (\BX \leftrightarrow \BY) \big]  \\
&\qquad \qquad \mp \frac{2}{g^2} \int \dt^d x \, \Tr \big[ \BX(C) \BY ( E^\pm C^{-1} )  -  (\BX \leftrightarrow \BY ) \big] , 
\end{split}
\end{align}
where
\begin{equation}
\begin{split}
\label{AE-defs}
{\hat A}_a^\pm &\equiv C \left( \lim_{r\to\pm\infty} |r|^{\frac{d}{2}-1} {\bar A}_a \right) C^{-1} , \\
E^\pm &\equiv \left( \lim_{u \mp \infty} \lim_{r \to \pm\infty} |r|^d \p_u {\bar A}_r  \right)  \pm \frac{1}{2} \int \dt u  \lim_{r \to \pm\infty} |r|^{d-2} \big[ {\bar A}^a , \p_u {\bar A}_a \big] . 
\end{split}
\end{equation}
Notice that even though the right-hand side of \eqref{symp_form} appears to depend on $\ci^\pm$, the boundary condition \eqref{matching} ensures that the symplectic form $\BOm$ does not.

To simplify this further, we invoke the results of \cite{He:2019jjk, He:2019pll, He:2019ywq}. In those papers, the authors decompose the gauge field into a \emph{radiative} part and a \emph{Coulombic} part so that
\begin{equation}
\begin{split}
\label{rad-coulomb-split}
    {\bar A}_\mu = {\bar A}_\mu^{R\pm} + {\bar A}_\mu^{C\pm} ,
\end{split}
\end{equation}
where the radiative part $(R)$ satisfies the free Maxwell's equations, whereas the Coulombic part $(C)$ is the inhomogeneous solution to Maxwell's equations and describes the self-interaction of the gauge field and, more generally, the interaction of the gauge field with the charged matter fields. This is defined by integrating the charge current against a Green's function. In the above equation, the $\pm$ in the superscript does not imply we are taking the $r \to \pm\infty$ limit, but rather distinguishes the Green's function that is used to extract the Coulombic part: $+$ for the advanced Green's function and $-$ for the retarded Green's function. This choice ensures that in the far future ${\bar A}_\mu^{C+} \to 0$ and in the far past ${\bar A}_\mu^{C-} \to 0$. Consequently, ${\bar A}_\mu^{R+}$ describes the outgoing radiative mode and ${\bar A}_\mu^{R-}$ the incoming radiative mode. In the quantum theory, they create the $out$ and $in$ one-particle gluon states respectively, and \cite{He:2019pll} showed that these gauge fields fall off at large $|r|$ as
\begin{align}
\label{large-r-falloffs-r}
{\bar A}_r^{R\pm} &= O ( |r|^{-\frac{d}{2}-1} ) + O ( |r|^{-d} )  , && {\bar A}_r^{C\pm} = O ( |r|^{-d} ) , \\
\label{large-r-falloffs-a}
{\bar A}_a^{R\pm} &= O ( |r|^{-\frac{d}{2}+1} ) + O(|r|^{-d+1} )  , && {\bar A}_a^{C\pm} =  O(|r|^{-d+1} ) . 
\end{align}
It is important to note that in \cite{He:2019pll}, the authors considered the large $r$ expansion of the gauge fields in a frame where $C=1$, so the expansions, in fact, apply to the gauge field ${\bar A}_\mu$, \emph{not} $A_\mu$.

We can now use \eqref{rad-coulomb-split} to simplify \eqref{AE-defs}. First, it is clear from \eqref{large-r-falloffs-a} that ${\hat A}^\pm_a$ is well-defined and only receives a contribution from the radiative part of the gauge field, as the Coulombic part falls off too quickly at large $|r|$. This means we can write
\begin{equation}
\begin{split}
\label{Adef-1}
{\hat A}_a^\pm &= C \Big(  \lim_{r\to\pm\infty} |r|^{\frac{d}{2}-1} {\bar A}^{R\pm}_a \Big) C^{-1} . 
\end{split}
\end{equation}
Next, we decompose $E^\pm$ into radiative modes, resulting in
\begin{equation}
\begin{split}
\label{Epm-explicit-1}
    E^\pm &= \Big( \lim_{u \mp \infty} \lim_{r \to \pm\infty} |r|^d \p_u {\bar A}^{R\pm}_r  \Big) \\
&\qquad \qquad + \Big( \lim_{u \mp \infty} \lim_{r \to \pm\infty} |r|^d \p_u {\bar A}^{C\pm}_r  \Big)   \pm \frac{1}{2} \int \dt u  \lim_{r \to \pm\infty} |r|^{d-2} \big[ {\bar A}^{R\pm a} , \p_u {\bar A}^{R\pm}_a \big] . 
\end{split}
\end{equation}
Note that in the last term, we have only kept the radiative contribution since the Coulombic part again falls off too quickly at large $|r|$. To further simplify this, we use (4.7) and (4.8) of \cite{He:2019pll}, which in our notation is given by\footnote{The exact form of (4.7), given (4.8), in \cite{He:2019pll} is
\begin{align*}
    2 \p_u F_{ur}^{(C\pm,d)} = \big[ A^{a(R\pm,\frac{d}{2}-1)} , F_{ua}^{(R\pm,\frac{d}{2}-1)} \big],
\end{align*}
where we dropped the matter current contribution since we are considering a pure Yang-Mills theory. The notation used in that paper translates to one used here as
\begin{align*}
    F_{ur}^{(C\pm,d)} = C \Big( \lim_{r \to \pm\infty} |r|^d \p_u {\bar A}_r^{C\pm} \Big) C^{-1} , \qquad A_a^{(R\pm,\frac{d}{2}-1)} = C \Big( \lim_{r \to \pm\infty} |r|^{\frac{d}{2}-1} {\bar A}_a^{R\pm} \Big) C^{-1}.
\end{align*}
}
\begin{equation}
\begin{split}
    2 \lim_{r \to \pm \infty} |r|^d \p_u^2 \bar A_r^{C\pm} = \lim_{r \to \pm \infty} |r|^{d-2} \big[ {\bar A}^{R\pm a} , \p_u {\bar A}^{R\pm}_a \big] .
\end{split}
\end{equation}
Integrating over $u$ and using the fact that $A_r^{C\pm}$ vanishes on $\CI^\pm_\pm$, we find
\begin{equation}
\begin{split}
\lim_{u \to \mp \infty} \lim_{r \to \pm \infty} |r|^d \p_u {\bar A}_r^{C\pm} = \mp \frac{1}{2} \int \dt u \lim_{r \to \pm \infty} |r|^{d-2} \big[ {\bar A}^{R\pm a} , \p_u {\bar A}^{R\pm}_a \big] .
\end{split}
\end{equation}
Using this, we find that the second line in \eqref{Epm-explicit-1} cancels exactly and we have
\begin{equation}
\begin{split}
\label{Epm-explicit}
E^\pm &= \lim_{u \mp \infty} \lim_{r \to \pm\infty} |r|^d \p_u {\bar A}^{R\pm}_r  . 
\end{split}
\end{equation}
Note that given \eqref{large-r-falloffs-r}, it seems like the limit here is divergent. However, it turns out that all these divergent terms die off at large $|u|$ \cite{He:2019jjk}, so the final result here is actually finite.

We can now evaluate the explicit form of the limit of $\bar A_a^{R\pm}$ in \eqref{Adef-1} and \eqref{Epm-explicit}. Since the radiative part satisfies the homogeneous Maxwell's equation, it admits a mode expansion. Thus, we can write
\begin{equation}
\begin{split}\label{mode-exp}
    {\bar A}^{R\pm}_\mu(X) = g \int \frac{\dt^{d+1}p}{(2\pi)^{d+1}} \frac{1}{2p^0} \ve_\mu^a(\vec{p}\,) \Big[ {\bar\CO }^\pm_a(\vec{p}\,) e^{ i p \cdot X} + \cc \Big] , \qquad p^\mu = ( |\vec{p}\,| , \vec{p}\,) , 
\end{split}
\end{equation}
where the polarization tensor satisfies $\ve_u^a(p) = 0$ (by our gauge condition) and $p^\mu \ve_\mu^a(\vec{p}\,) = 0$ (by transversality condition). We can now perform a large $r$ expansion of \eqref{mode-exp} to extract the leading large $r$ term and determine both $\hat A_a^\pm$ in \eqref{Adef-1} and $E^\pm$ in \eqref{Epm-explicit}. This involves changing the integration variable from $\vec{p}$ to $(\o,y^a)$, so that the momentum vector and polarization tensor are now written as
\begin{equation}
\begin{split}
\label{mompar}
\vec{p}(\o,y) = \o \left( y^a , \frac{1-y^2}{2} \right) , \qquad \ve_\mu^a(\vec p\,) = ( - y^a, \d^a_b , - y^a ) .
\end{split}
\end{equation}
We can then use the stationary phase approximation to localize the integral over $y$ to $x$ in the large $|r|$ limit. This process is explained in detail in \cite{He:2019ywq, He:2019pll, He:2019jjk}, and the limit in \eqref{Adef-1} simplifies to
\begin{equation}
\begin{split}\label{Abar-def}
{\hat A}^\pm_a (u,x) &= \pm \frac{g}{2(2\pi)^{\frac{d}{2}+1}}  \int_0^\infty \dt \o \,\o^{\frac{d}{2}-1} \left[ \CO^\pm_a(\o,x)  e^{-\frac{i}{2}\o u \mp \frac{i\pi d}{4} } + \cc  \right] . 
\end{split}
\end{equation}
where
\begin{equation}
\begin{split}
\label{Oa-def}
\CO_a^\pm(\o,x) \equiv C {\bar \CO}_a^\pm(\o,x) C^{-1} . 
\end{split}
\end{equation}
Once we quantize the theory in Section~\ref{sec:quantization}, we will see that for $\w > 0$, $\CO_a^\pm(\w,x)$ is the annihilation operator of a gluon with polarization $a$ and momentum ${\vec p}(\w,x)$, while $\CO_a^\pm(\w,x)^\dag$ is the corresponding creation operator.

The limit in \eqref{Epm-explicit} is more complicated to evaluate and requires one to consider the large $r$ expansion separately for $d$ odd and even.\footnote{It can be shown that the mode expansion can be recast in terms of Bessel functions $K_\nu(\sqrt{iu}/\sqrt{ir})$ with $\nu=\frac{d}{2}-1+\mzz$. The large $r$ expansion then corresponds to the asymptotic expansion of $K_\nu(z)$ at small $z$. This is qualitatively different depending on whether $\nu$ is an integer or half-integer.} This is done explicitly in \cite{He:2019jjk}, where it was determined that
\begin{align}\label{Edef}
\begin{split}
    E^\pm(x) = \begin{cases}
\pm  \frac{g^2}{2(4\pi)^{\frac{d}{2}} \G\left(\frac{d}{2}\right)} (-\p^2)^{\frac{d}{2}-1} \p^a \Nfield_a^\pm(x)  & \text{$d$ even} \\
\pm \frac{g^2 (-1)^{\frac{d-1}{2}} \G(d-1) }{8\pi^{d+1}} \displaystyle\int \dt^dy \frac{\p^a \Nfield_a^\pm(y)}{\left[ (x-y)^2 \right]^{d-1}} & \text{$d$ odd} ,
\end{cases}
\end{split}
\end{align}
where\footnote{A similar soft operator, which we denote as $\hat N^\pm_a$, was defined in (3.75) of \cite{He:2020ifr}, and it is related to the one defined in \eqref{Nadef} via $ \hat N_a^\pm = - \frac{g^2}{4\sqrt 2} C \Nfield_a^\pm C^{-1}$.}
\begin{equation}
\begin{split}
\label{Nadef}
\Nfield^\pm_a(x) \equiv \frac{1}{g} \lim_{\o \to 0}  [ \o {\bar \CO}^\pm_a(\o,x)  ] . 
\end{split}
\end{equation}
Thus, $\Nfield_a^\pm$ creates a soft gluon with polarization $a$. Furthermore, it was shown in \cite{He:2019ywq} that this operator is Hermitian
\begin{equation}
\begin{split}
\label{Na-hermitian}
\Nfield^{\pm}_a(x)^\dagger= \Nfield^{\pm}_a(x)
\end{split}
\end{equation}
and satisfies a flatness constraint
\begin{align}
\begin{split}
\label{Na-flat}
\p_{[a} \Nfield^\pm_{b]}(x) = 0 \quad \implies \quad \Nfield^\pm_a(x) = \p_a \Nfield^\pm(x) .
\end{split}
\end{align}
The Hermiticity constraint arises from the requirement that the asymptotic expansion of the gauge field is analytic, and the flatness constraint arises from the requirement that the symplectic form is invertible, as was discussed in \cite{He:2020ifr}.\footnote{In \cite{He:2019ywq}, the operator $N^\pm_a(x)$ was called $\CO_a^{(\pm,0)}(x)$, and the Hermiticity and flatness conditions are discussed in equations (2.21) and (5.14) of that paper, respectively.\label{fn:newN-oldN}}

We see from \eqref{Edef} that there is a qualitatively different structure in odd and even dimensions. The relationship between $E^\pm$ and $\Nfield^\pm_a$ is local in even dimensions but involves a non-local integral over $y$ in odd dimensions. This is due to the fact that Huygens's principle is satisfied in even dimensions but not in odd dimensions. More precisely, the Green's function in even dimensions is localized on the light cone whereas in odd dimensions it has support everywhere in the interior of the light cone. Nevertheless, despite these different structures in odd and even dimensions, it is possible to unify them via a \emph{shadow transform}. This is an integral transform that maps a conformal primary operator (in a CFT) of scaling dimension $\D$ to another primary operator with dimension $d-\D$. For vector primaries, this is defined as
\begin{equation}
\begin{split}
\label{shadow_def}
{\wt V}_a(x) \equiv \int \dt^d y \frac{ \CI_{ab}(x-y) }{ [ ( x - y )^2 ]^{d-\D}} V^b(y) , \qquad \CI_{ab}(x) = \d_{ab} - 2 \frac{x_ax_b}{x^2}. 
\end{split}
\end{equation}
Up to a normalization factor, the shadow transform is its own inverse, i.e.
\begin{equation}\label{double-shadow}
\begin{split}
\wt{\wt{V_a}}(x) = c_{\D,1} V_a(x) , \qquad c_{\D,1} = \frac{\pi^d(\D-1)(d-\D-1) \G(\frac{d}{2}-\D)\G(\D-\frac{d}{2})}{\G(\D+1)\G(d-\D+1)} . 
\end{split}
\end{equation}
Using the shadow transform, we prove in Appendix~\ref{app:shadow} that both cases of \eqref{Edef} can be unified via
\begin{align}\label{Edef_final}
E^\pm = \pm \frac{g^2}{4 c_{1,1}} \p^2 \CN^\pm , \qquad \p_a \CN^\pm = \wt{\p_a \Nfield^\pm} . 
\end{align}
Substituting \eqref{Abar-def} and \eqref{Edef_final} into \eqref{symp_form}, the full symplectic form becomes
\begin{align}
\begin{split}
\label{symp_form_final}
\BOm(\BX,\BY) &= \frac{i}{2(2\pi)^{d+1}} \int \dt^d x  \int_0^\infty \dt \o \, \o^{d-1} \, \Tr \left[   \delta^{ab}\BX( \CO^\pm_a) \BY( \CO_b^{\pm\dagger}) - (\BX \leftrightarrow \BY) \right] \\
&\qquad \qquad \qquad \qquad \qquad \qquad - \frac{1}{2 c_{1,1}} \int \dt^d x \, \Tr \big[ \BX(C) \BY (  \p^2 \CN^\pm C^{-1}  )  -  (\BX \leftrightarrow \BY ) \big] .
\end{split}
\end{align}

\section{Asymptotic Hilbert Space}
\label{sec:quantization}

In the previous section, we have restricted ourselves to a purely classical analysis. We now promote the classical fields to quantum fields via canonical quantization. In Section~\ref{ssec:commutators}, we extract the quantum commutators from the symplectic form.
Then, in Section~\ref{ssec:hilbert}, we construct the full Hilbert space involving the gauge field, including the vacuum sector.

\subsection{Quantum Commutators}\label{ssec:commutators}

By inverting the symplectic form \eqref{symp_form_final}, we can determine the quantum commutators of the operators. To do this, we expand out the Lie algebra indices in \eqref{symp_form_final} in the adjoint representation. Since $\CN^\pm, \CO^\pm_a \in \mfg$, we can write
\begin{equation}
\begin{split}
\CO^\pm_a = \CO^{\pm I}_a T^I , \qquad \CN^\pm = \CN^{\pm I} T^I , \qquad (T^I)^{JK} = - f^{IJK} . 
\end{split}
\end{equation}
Furthermore, since $C \in \CG$, it has indices $C^{IJ}$, and its inverse is given by $(C^{-1})^{IJ} = ( C^T )^{IJ}  = C^{JI}$.\footnote{In the adjoint representation, $(T^I)^T = - T^I$, and so $(e^{\a^I T^I})^{-1}=e^{-\a^IT^I} = e^{\a^I(T^I)^T} = (e^{\a^I T^I })^T$.} It also satisfies the identity\footnote{The general identity that applies to any representation $R_i$ of $\CG$ is $R_i(C)^{-1} R_i(T^I) R_i(C) = C^{IJ} R_i(T^J)$. Note then that \eqref{C_cube_identity} is the special case of this applied to the adjoint representation.}
\begin{equation}
\begin{split}\label{C_cube_identity}
f^{I'J'K'}C^{II'}C^{JJ'}C^{KK'} = f^{I'J'K'}C^{I'I}C^{J'J}C^{K'K}  = f^{IJK} .
\end{split}
\end{equation}
Using these relations, we can explicitly take the trace of \eqref{symp_form_final}, so that it becomes
\begin{align}
\begin{split}
\label{symp_form_final_1}
\BOm(\BX,\BY) &= \frac{ i}{2(2\pi)^{d+1}} \int \dt^d x  \int_0^\infty \dt \o \,\o^{d-1}  \delta^{ab} \left(  \BX( \CO^{\pm I}_a) \BY( \CO_b^{\pm I\dagger}) - (\BX \leftrightarrow \BY ) \right) \\
&\qquad \qquad \qquad - \frac{1}{2 c_{1,1}} f^{JKL} \int \dt^d x  \bigg(  \BX( C^{IJ} ) \BY (  \p^2 \CN^{\pm K} C^{IL} )  - ( \BX \leftrightarrow \BY ) \bigg)  . 
\end{split}
\end{align}

With everything written out explicitly, we can now invert this following the procedure given in Section 3.3.6 of \cite{He:2020ifr} to obtain the Dirac brackets, which then give rise to the quantum commutators (see Footnote~\ref{fn:dirac-comm})
\begin{align}
\label{db_1}
\left[ \CO_a^{\pm I}(\o,x)  , \CO_b^{\pm J}(\o',y)^\dag \right] &= \frac{2(2\pi)^{d+1}}{\w^{d-1}} \d_{ab} \d^{IJ} \d(\o-\o') \d^d(x-y) \\
\label{db_2}
\left[ C^{IJ}(x), C^{KL}(y) \right] &= 0 \\
\label{db_3}
\left[ \CN^{\pm I}(x), C^{JK}(y) \right] &= 2 i c_{1,1} f^{IKL} C^{JL}(y) G(x-y) \\
\label{db_4}
\left[ \CN^{\pm I}(x), \CN^{\pm J}(y) \right] &= 2 i c_{1,1} f^{IJK} \int \dt^d z \, G(x-z) G(y-z) \p^2 \CN^{\pm K}(z),
\end{align}
where $G(x)$ is the scalar Green's function satisfying $\p^2 G(x) = \d^d(x)$ and is explicitly given by
\begin{equation}
\begin{split}
G(x) = \begin{cases}
\frac{1}{4\pi} \ln(x^2) , & d = 2,  \\
- \frac{\G\left(\frac{d}{2}-1\right)}{4\pi^{\frac{d}{2}}} \frac{1}{(x^2)^{\frac{d}{2}-1}}  , & d > 2 . 
\end{cases}
\end{split}
\end{equation}
Now, since the momentum $\vec{p}(\o,x)$ in flat null coordinates is parametrized via \eqref{mompar}, it can be deduced that
\begin{equation}
\begin{split}
    \frac{2}{\w^{d-1}} \d(\o-\o') \d^d(x-x') = 2 | \vec{p}\,|\,  \d^{d+1} ( \vec{p} - \vec{p}\,') .
\end{split}
\end{equation}
Consequently, \eqref{db_1} is precisely the commutation relation for a pair of creation and annihilation operators, and so we interpret $\CO_a^{\pm I}(\o,x)$ as an operator that annihilates (by acting on the vacuum state) an outgoing $(+)$ or incoming $(-)$ one-particle gluon state with color $I$, polarization $a$, and momentum $\vec{p}(\o,x)$.

\subsection{Canonical Quantization}\label{ssec:hilbert}

Having derived the quantum commutators in the previous section, we now proceed to canonically quantize the theory. We start with the vacuum sector of the theory, which is spanned by the operators $C$ and $\CN^\pm$. Since $C$ commutes with itself, we can work in a basis of states $|U,\pm \ra$ that diagonalizes this operator:
\begin{equation}
\begin{split}
\label{U_def}
C(x) \ket{U,\pm} =  U(x) \ket{U,\pm} , \qquad \braket{ U,\pm }{ U', \pm } = \d(U-U') . 
\end{split}
\end{equation}
As discussed in \cite{He:2020ifr}, the state with $U(x)=1$ is Lorentz invariant, and more general vacuum states are given by
\begin{equation}
\begin{split}
\label{fvac_exp}
    \ket{f,\pm} = \int [ \msD U ] \, f(U) \ket{U,\pm} . 
\end{split}
\end{equation}
To complete the description of the vacuum Hilbert space, we determine after some algebra the action of $\CN^\pm$ on the basis states via \eqref{db_3} to be
\begin{align}
\begin{split}
\label{N_action}
\CN^{\pm I}(x) \ket{U,\pm} = 2i c_{1,1} \int \dt^d y\, G(x-y) U^{JI}(y) \mfD^J_{U(y)}\ket{U,\pm} , 
\end{split}
\end{align}
where $\mfd^I_U$ is a derivative operator defined by
\begin{equation}
\begin{split}
\mfd^I_{U(x)} U(y) = - T^I U(x) \d^d(x-y) . 
\end{split}
\end{equation}
This operator was introduced and extensively studied in Appendix B of \cite{He:2020ifr}, and we refer the reader there for more details regarding its explicit form and properties. It can then be easily verified that \eqref{N_action} is consistent with \eqref{db_4}.

With the vacuum sector fully characterized, the rest of the Hilbert space is constructed as follows. The annihilation operators by definition annihilate all the vacuum states, so that
\begin{equation}
\begin{split}
\label{O_annihilation}
\CO_a^{\pm I} (\o,x)  \ket{U,\pm} = 0 .
\end{split}
\end{equation}
The remaining states are then constructed as a Fock space by repeatedly acting on the vacuum states with the creation operators $\CO^{\pm I}_a(\o,x)^\dag$. This completes our description of the full Hilbert space.

Analogous to the four-dimensional case considered in \cite{He:2020ifr}, the charge that generates LGTs is from \eqref{large_gauge_charge} 
\begin{align}
\begin{split}
\label{large_gauge_charge_1}
Q_\ve &=  \pm \frac{1}{g^2} \oint_{\ci^\pm_\mp} \Tr [ \ve \star \msF ] \\
&= - \frac{1}{2c_{1,1}} \int \dt^d x\, \ve^I(x) C^{ IJ}(x) \p^2 \CN^{\pm J}(x) - \frac{1}{g^2} \int \dt u\, \dt^d x\, \Tr \left[ \ve(x)  \big[ {\hat A}^{\pm a}(u,x) , \p_u {\hat A}^{\pm}_a (u,x) \big] \right] , 
\end{split}
\end{align}
where the commutator in the second term is the Lie algebra commutator and not the quantum commutator. Notice that $Q_\ve$ does not require a $\pm$ superscript since it is constructed from the symplectic form, which itself is equal on $\ci^+$ and $\ci^-$ due to the boundary condition we imposed in \eqref{C-def}. Using \eqref{Abar-def}, the second term can be rewritten as
\begin{equation}
\begin{split}
&\frac{1}{g^2} \int \dt u \, \dt^d x \, \Tr \left[ \ve(x) \big[ {\hat A}^{\pm a}(u,x) , \p_u {\hat A}^{\pm}_a(u,x) \big] \right] \\
&\qquad =  \frac{i f^{IJK} \d^{ab}}{2(2\pi)^{d+1}} \int_0^\infty \dt \o \, \o^{d-1} \int \dt^d x \, \ve^I(x) \CO^{\pm J}_a(\o,x)^\dagger \CO^{\pm K}_b(\o,x)  .
\end{split} 
\end{equation}
Using this and the commutators \eqref{db_1}--\eqref{db_4}, it is easy to check that $[Q_\ve , \cdot ] = - i \d_\ve ( \cdot )$, so that
\begin{align}\label{Q-gauge-comm}
    \big[ Q_\ve , \CO^{\pm I}_a(\o,x) \big] &=  - i f^{IJK} \ve^J(x)  \CO^{\pm K}_a(\o,x)  \\
    \big[ Q_\ve , \CO^{\pm I}_a(\o,x)^\dag \big] &=  - i f^{IJK} \ve^J(x)  \CO^{\pm K}_a(\o,x)^\dag  \\
    \big[ Q_\ve , C(x)\big] &= -i \ve(x) C(x) \\
    \big[Q_\ve , \CN^\pm (x) \big] &= 0.
\end{align}
Lastly, by \eqref{N_action} and \eqref{O_annihilation}, we find
\begin{align}
\begin{split}
\label{charge_action}
Q_\ve \ket{U,\pm} &= -i \int \dt^d x\, \ve^I(x) \mfD^I_{U(x)} \ket{U,\pm} .
\end{split}
\end{align}

\section{Ward Identity and the Leading Soft Gluon Theorem}
\label{sec:wardid}

Thus far, our discussion has been limited to pure Yang-Mills theory. However, for the rest of the paper, we will generalize our analysis to include massless states that transform under arbitrary representations $R_i$ of $\CG$. In Section~\ref{ssec:ward}, we will derive the Ward identity associated with LGTs. We then show that the Ward identity implies the leading soft gluon theorem in Section~\ref{ssec:soft_thm}. 

\subsection{Ward Identity} \label{ssec:ward}

Consider an $n$-point scattering amplitude,\footnote{More generally, we can consider scattering amplitudes with arbitrary $in$ and $out$-vacuum states given in \eqref{fvac_exp}, but they can all be described in terms of \eqref{S_matrix}.}
\begin{align}
\label{S_matrix}
\bra{ U,+ } T \left\{ \CO_1  \cdots \CO_n \right\} \ket{ U' , - } ,
\end{align}
where $T$ is the time-ordering operator and $\CO_i \equiv \CO_i(\o_i,x_i)$ is defined by\footnote{Comparing this definition of $\CO_i$ to that given in Section 4.3 of \cite{He:2020ifr}, we see that \eqref{Oins_def} has an extra factor of $R_i(C(x_i))$. This is because $\bar \CO_i^\pm$ are gauge-invariant operators, whereas $\CO_i^\pm$ defined in \cite{He:2020ifr} are gauge-covariant operators, and the relationship between them is captured via multiplication by $R_i(C(x_i))$.}
\begin{equation}
\begin{split}
\label{Oins_def}
\CO_i &\equiv R_i( C(x_i) ) \left( \t(\o_i)  \left[ {\bar\CO}_i^{+}(\o_i,x_i) - {\bar\CO}_i^{-}(\o_i,x_i)  \right] + \t(-\o_i ) \left[ {\bar\CO}_i^{-}(-\o_i,x_i)^\dag - {\bar\CO}_i^{+}(-\o_i,x_i)^\dag \right] \right) ,
\end{split}
\end{equation}
where $\theta$ is the Heaviside step function. These operators in the scattering amplitude appear rather complicated, but for hard operators \eqref{Oins_def} is rather simple. For instance, consider $\o_i > 0$. In this case, only the first term in \eqref{Oins_def} is non-zero. When inserted into \eqref{S_matrix}, the time-ordering operator moves ${\bar\CO}^-$ to the right where it annihilates the $in$-vacuum. Consequently, if $\w_i>0$, we have $\CO_i = R_i( C(x_i) ) {\bar\CO}_i^{+}(\o_i,x_i) $. Likewise, a similar argument shows that for $\w_i < 0$ we have $\CO_i =  R_i( C(x_i) ) {\bar\CO}_i^{-}(-\o_i,x_i)^\dag$. To summarize, we have
\begin{align}
    \CO_i(\o,x)  = \begin{cases}
        R_i(C(x)) \bar \CO_i^+(\w,x), & \w > 0 \\
        R_i(C(x)) \bar \CO^-_i(-\w,x)^\dag, & \w < 0.
    \end{cases}
\end{align}
The complicated expression \eqref{Oins_def} is necessary only for the case $\o_i=0$. For instance, the soft gluon operator \cite{Kapec:2017gsg,Kapec:2021eug} is defined by
\begin{equation}
\begin{split}
\label{Sadef}
S^I_a(x) \equiv \lim_{\o \to 0} \o \CO^I_a(\o,x) = C^{IJ}(x) \big[ N_a^{+J} (x) -  N_a^{-J}(x) \big] ,
\end{split}
\end{equation}
where we have used \eqref{Nadef} and \eqref{Na-hermitian}, as well as the regulated Heaviside step function $\theta(0) = \frac{1}{2}$, for the second equality. 

To derive the Ward identity associated with LGTs, first note that $C(x)$ commutes with all operators with energies not strictly zero. Hence, we have
\begin{align}
\la U,+| C(x) T\{ \CO_1 \cdots \CO_n \}|U',- \ra = \la U,+| T\{ \CO_1 \cdots \CO_n \} C(x)|U',-\ra.
\end{align}
Because the vacua are $C$ eigenstates, it follows that
\begin{align}
( U(x) - U'(x) ) \la U,+| T\{ \CO_1 \cdots \CO_n \} |U',- \ra = 0,
\end{align}
which immediately implies
\begin{align}\label{amp1}
\la U,+| T\{ \CO_1 \cdots \CO_n \} |U',- \ra = \delta(U-U') \la \CO_1 \cdots \CO_n \ra_U,
\end{align}
where the correlator is implicitly time-ordered.

Next, recall that the large gauge charge $Q_\ve$ given in \eqref{large_gauge_charge} generates LGTs (see \eqref{Q-gauge-comm}). This means the action of $Q_\ve$ on an arbitrary field $\CO_i$ in representation $R_i$ is given by
\begin{align}
    \big[Q_\ve, \CO_i \big] = -i R_i(\ve(x_i)) \CO_i.
\end{align}
It follows that when inserting $Q_\ve$ on the left of all the hard operators $\CO_i$ and then commuting it past all the $\CO_i$'s, we obtain
\begin{align}
\begin{split}
\bra{ U,+ } \big[ Q_\ve ,  T\{\CO_1 \cdots \CO_n\} \big] \ket{ U',- } = - i \sum_{i=1}^n R_i(\ve(x_i)) \bra{ U,+ } T\{\CO_1 \cdots \CO_n\} \ket{ U',- }  .
\end{split}
\end{align}
Using \eqref{charge_action}, the above equality becomes
\begin{align}
\begin{split}
&\int \dt^d y \, \ve^I(y) \big( \mfD^I_{U(y)} + \mfD^I_{U'(y)} \big)  \bra{ U,+ } T\{\CO_1 \cdots \CO_n\} \ket{ U',- } \\
&\qquad \qquad \qquad \qquad \qquad = -  \sum_{i=1}^n R_i(\ve(x_i)) \bra{ U,+ } T\{\CO_1 \cdots \CO_n\} \ket{ U',- } .
\end{split}
\end{align}
Substituting in \eqref{amp1} and using the property $\big( \mfD^I_{U(y)} + \mfD^I_{U'(y)}\big) \d(U-U') = 0$ derived in \cite{He:2020ifr}, we find 
\begin{align}
\begin{split}
 \int \dt^d y\, \ve^I(y) \mfD^I_{U(y)}  \la \CO_1 \cdots \CO_n \ra_U  = - \sum_{i=1}^n R_i(\ve(x_i))  \la \CO_1 \cdots \CO_n \ra_U . 
\end{split}
\end{align}
Finally, setting $\ve(y) = T^I \d^d(x-y)$, we obtain the differential equation
\begin{equation}
\begin{split}
\label{Du_identity}
\mfd^I_{U(x)} \avg{ \CO_1\cdots \CO_n  }_U = - \sum_{i=1}^n \d^d ( x - x_i ) R_i(T^I) \avg{ \CO_1\cdots \CO_n  }_U , 
\end{split}
\end{equation}
This is a first-order differential equation, which we can solve to get the Ward identity
\begin{align}
\label{ward_id}
\avg{ \CO_1\cdots \CO_n  }_U = R_1(U(x_1))  \cdots R_n(U(x_n)) \avg{ \CO_1\cdots \CO_n  }_{U=1},
\end{align}
where $\avg{ \CO_1\cdots \CO_n  }_{U=1}$ is an integration constant, and is the scattering amplitude evaluated in standard QFT. This is the main result of this section, and we will use it to derive the leading soft gluon theorem in the next subsection.

\subsection{Leading Soft Gluon Theorem}\label{ssec:soft_thm}

The leading soft gluon theorem states that the insertion of the soft gluon operator \eqref{Sadef} is given by
\begin{equation}
\begin{split}
\label{soft_thm}
\avg{ S_a^I(x) \CO_1\cdots \CO_n  }_{U=1} &= i g \sum_{i=1}^n \o \frac{p_i \cdot \ve_a (x) }{ p_i \cdot p } R_i(T^I) \avg{ \CO_1\cdots \CO_n  }_{U=1} \\
&= i g \sum_{i=1}^n \p_a \ln \big[ ( x - x_i )^2 \big] R_i(T^I) \avg{ \CO_1 \cdots \CO_n }_{U=1} ,
\end{split}
\end{equation}
where in the second equality, we have used the explicit parameterization \eqref{mompar} for both the momenta and polarization. In this subsection, we derive this from \eqref{ward_id} (or equivalently \eqref{Du_identity}).

Let us start with the left-hand side of the above equation along with the definition \eqref{Sadef}. We need to determine the action of $\Nfield_a^\pm(x)$ on the vacuum state. Recalling \eqref{double-shadow}, we can easily invert \eqref{Edef_final} so that
\begin{equation}
\begin{split}
\Nfield_a^{\pm I} (x) = \frac{1}{c_{1,1}} \wt{\p_a \CN^{\pm I}}(x) . 
\end{split}
\end{equation}
Using the fact that $\p_a G(x)$ has conformal dimension $\D=d-1$, we can evaluate its shadow transform \eqref{shadow_def} to be
\begin{equation}
\begin{split}
\wt{\p_a G}(x) = \frac{1}{2} \p_a \ln(x^2).
\end{split}
\end{equation}
It then follows from this and \eqref{N_action} that the action of $N^\pm_a$ on the vacuum state is given by
\begin{align}\label{Naction2}
\begin{split}
\Nfield_a^{\pm I} (x) \ket{U,\pm} = i \int \dt^d y \p_a \ln\big[(x-y)^2\big] U^{JI}(y) \mfD^J_{U(y)}\ket{U,\pm}.
\end{split}
\end{align}
Hence, upon inserting $S_a^I(x)$ between any two vacua $U$ and $U'$, we get
\begin{align}
\begin{split}
&\la U,+| T\{ S_a^I(x) \CO_1\cdots \CO_n \} |U',- \ra \\
&= \la U,+| C^{IJ}(x) T\{ (N_a^{J+}(x) - N_a^{J-}(x) )\CO_1\cdots \CO_n \} |U',- \ra \\
&= -i g U^{IJ}(x) \int \dt^d y\, \p_a \ln \big[ (x-y)^2\big] \big( U^{KJ}(y) \mfD_{U(y)}^K + U'^{KJ}(y) \mfD_{U'(y)}^K \big)\la U,+| T\{ \CO_1 \cdots \CO_n \} | U',-\ra ,
\end{split}
\end{align}
where the time-ordering operator $T$ moves $N_a^{+}$ to the left to act on the $out$-vacuum and $N_a^-$ to the right to act on the $in$-vacuum. Substituting in \eqref{amp1} on both sides of the above equation and simplifying, we arrive at the expression
\begin{equation}
\begin{split}
\avg{ S^I_a(x) \CO_1 \cdots \CO_n }_U &= - i g U^{IJ}(x) \int \dt^d y \, \p_a \ln\big[(x-y)^2\big] U^{KJ}(y) \mfD^K_{U(y)} \avg{ \CO_1 \cdots \CO_n }_U .
\end{split}
\end{equation}
Finally, substituting in \eqref{Du_identity}, we get
\begin{align}\label{soft-general}
    \avg{ S^I_a(x) \CO_1 \cdots \CO_n }_U &= ig U^{IJ}(x) \sum_{i=1}^n \p_a \ln \big[ (x-x_k)^2 \big] U^{KJ}(x_k) R_i(T^K) \avg{ \CO_1 \cdots \CO_n }_U,
\end{align}
and then setting $U=1$, we reproduce the leading soft gluon theorem \eqref{soft_thm}. Of course, the result we have just derived is more general than \eqref{soft_thm}, since it can be evaluated for any $U$ whereas the soft theorem applies only for $U=1$. However, we can also consider scattering amplitudes with \emph{multiple} soft gluons where the soft limits are taken consecutively.\footnote{The symmetry interpretation of simultaneous soft limits is still an open problem.} It can be shown that this general multiple soft gluon theorem is in fact completely equivalent to \eqref{soft-general}.

\section*{Acknowledgements} 

We would like to thank Daniel Kapec for useful conversations that initiated this work. T.H. has been supported by the Heising-Simons Foundation ``Observational Signatures of Quantum Gravity'' collaboration grant 2021-2817, the U.S. Department of Energy grant DE-SC0011632, and the Walter Burke Institute for Theoretical Physics. P.M. gratefully acknowledges support from the STFC consolidated grants ST/P000681/1 and ST/T000694/1.

\appendix

\section{Definitions and Conventions}\label{app:conventions}

In this appendix, we provide a lightning overview of the conventions we use from differential geometry and Lie algebra. We will be following the conventions outlined in \cite{He:2020ifr}, in which more details can be found (see also Chapter 20 of \cite{Nair:2005iw}). 

\subsection{Differential Geometry} 

We will mainly be interested in two manifolds -- the phase space symplectic manifold and the spacetime Lorentzian manifold. Boldfaced letters are used to denote forms and vectors on the phase space manifold, and regular letters denote those on the spacetime manifold.

\subsubsection{Spacetime Manifold}
\label{spacetime-conventions}

Let our spacetime $\CM$ to be a $(d+2)$-dimensional globally hyperbolic Lorentzian manifold with coordinates $x^\mu$ and metric $g_{\mu\nu}$, where $\mu,\nu \in\{ 0,1,\ldots,d+1\}$. At every point $p \in \CM$, we adopt the canonical basis $\p_\mu$ on the tangent space $T_p\CM$ and $\dt x^\mu$ on the cotangent space $T^*_p\CM$. The wedge product is defined as
\begin{align}
\dt x^{\mu_1} \wedge \cdots \wedge \dt x^{\mu_q} \equiv q! \, \dt x^{[\mu_1} \otimes \cdots \otimes \dt x^{\mu_q]},
\end{align}
where $[\,\cdots]$ denote the weighted antisymmetrization of indices, e.g. $\w^{[\mu\nu]} = \frac{1}{2!} (\w^{\mu\nu} - \w^{\nu\mu})$. This is a basis on the space of $q$-forms $\O^q(\CM)$, which means any $q$-form $C_q$ can be written as
\begin{align}
    C_q = \frac{1}{q!}(C_q)_{\mu_1\cdots \mu_q}(x) \, \dt x^{\mu_1} \wedge \cdots \wedge \dt x^{\mu_q}, \quad C_q \in \O^q(\CM).
\end{align}
This implies that for any $p$-form $C_p$ and $q$-form $C_q$, we have
\begin{align}
(C_p \wedge C_q)_{\mu_1 \cdots \mu_{p+q}} = \frac{(p+q)!}{p!q!} (C_p)_{[\mu_1 \cdots \mu_p} (C_q)_{\mu_{p+1} \cdots \mu_{p+q}]}.
\end{align}
The volume form on $\CM$ is a $(d+2)$-form defined as
\begin{align}
\e \equiv \sqrt{-\det g}\, \dt x^0 \wedge \cdots \wedge \dt x^{d+1},
\end{align}
and the choice of the sign above fixes an orientation for the spacetime. It obeys the useful identity
\begin{align}
\e_{\mu_1\cdots \mu_p \a_{p+1}\cdots\a_{d+2}}\e^{\a_{p+1} \cdots \a_{d+2} \nu_1 \cdots \nu_p} = (-1)^s p! (d+2-p)! \d_{[\mu_1}^{\nu_1} \cdots \d_{\mu_p]}^{\nu_p},
\end{align}
where $s$ is the number of negative eigenvalues in the metric, as well as
\begin{align}
\e_{\mu_1\cdots \mu_{d+2}}\p^{\mu_1} C_{d+1}^{\mu_2\cdots\mu_{d+2}} = \n^{\mu_1} ( \e_{\mu_1\cdots\mu_{d+2}} C_{d+1}^{\mu_2\cdots\mu_{d+2}} ) 
\end{align}
for any $(d+1)$-form $C_{d+1}$. 

The following conventions are used for standard operations on forms, with $C_q$ being a $q$-form:
\begin{align}
\begin{split}
(i_\xi C_q)_{\mu_1 \cdots\mu_{q-1}} &\equiv \xi^\mu (C_q)_{\mu\mu_1\cdots \mu_{q-1}} \\
(\dt C_q)_{\mu_1\cdots\mu_{q+1}} &\equiv (q+1)\p_{[\mu_1} (C_q)_{\mu_2 \cdots \mu_{q+1}]} \\
(\star C_q)_{\mu_1 \cdots \mu_{d+2-q}} &\equiv \frac{1}{q!} \e_{\mu_1 \cdots \mu_{d+2-q}}{}^{\nu_1 \cdots\nu_q} (C_q)_{\nu_1 \cdots \nu_q}.
\end{split}
\end{align}
The exterior derivatives are nilpotent, i.e. $\dt^2 = 0$, which implies that all exact forms ($C_q=\dt C_{q-1}$) are closed ($\dt C_q = 0$). We can also verify
\begin{align}
\star^2 C_q = (-1)^{s+q(d+2-q)} C_q ,
\end{align}
where again $s$ is the number of negative eigenvalues in the metric.

Given a $q$-form, we can integrate it over a $q$-dimensional submanifold of $\CM$. In particular, we are mainly concerned with the cases $q = d,d+1$ and $d+2$, for which
\begin{align}\label{form-int}
\begin{split}
    \int_{\S_{d+2}} C_{d+2} &=(-1)^s \int_{\S_{d+2}} \dt^{d+2} x \sqrt{|g|} (\star C_{d+2})  \\
    \int_{\S_{d+1}} C_{d+1} &= -\int_{\S_{d+1}} \dt \S_\mu(\star C_{d+1})^\mu \\
    \int_{\S_{d}} C_{d} &= -\frac{1}{2} \int_{\S_{d}} \dt S_{\mu\nu} (\star C_{d})^{\mu\nu},
\end{split}
\end{align}
where $\dt \S_{\mu}$ and $\dt S_{\mu\nu}$ are respectively the directed area elements corresponding to the submanifolds $\S_{d+1}$ and $\S_{d}$. Stokes' theorem for a $q$-form is given by
\begin{align}
\label{Stokes_theorem}
\int_{\S_q} \dt C_{q-1} = \oint_{\p\S_q} C_{q-1},
\end{align}
where we orient $\p\S_q$ to be outwardly-directed with respect to $\S_q$. It is useful for us to focus on the special cases where $q=d+2$ and $q=d+1$, in which case Stokes' theorem is given to be
\begin{align}
\int_{\S_{d+2}} \e \n_\mu C^\mu = \oint_{\p\S_{d+2}} \dt \S_\mu C^\mu , \qquad \int_{\S_{d+1}} \dt \S_\mu \n_{\nu} C^{[\mu\nu]} = \frac{1}{2} \oint_{\partial\S_{d+1}} \dt S_{\mu\nu} C^{\mu\nu},
\end{align}
where $\n_\mu$ is the covariant derivative with respect to the metric $g$.

One of the (slight) challenges pertaining to working with forms is settling the explicit signs that determine the orientation of various area elements. We therefore discuss the signs that will be relevant for our work here. Flat null coordinates are related to Cartesian coordinates by
\begin{equation}
\begin{split}
X^\mu = \frac{r}{2} \left( 1 + x^2 + \frac{u}{r} , 2 x^a , 1 - x^2 - \frac{u}{r} \right)  .
\end{split}
\end{equation}
The oriented volume form is
\begin{equation}
\begin{split}
\e = \dt X^0 \wedge \cdots \wedge \dt X^{d+1} = \frac{1}{2} r^d \, \dt u \wedge \dt^d x \wedge \dt r  . 
\end{split}
\end{equation}
Given this and recalling the first equation in \eqref{form-int}, the integral over any $(d+2)$-form is given by
\begin{equation}
\begin{split}\label{D-integral}
\int_\CM C_{d+2} &= - \frac{1}{2} \int \dt u \, \dt r \,\dt^d x \,|r|^d ( \star C_{d+2} ) . 
\end{split}
\end{equation}

The area elements on $\ci^\pm$ and $\ci^\pm_\mp$ can be determined by requiring that Stokes' theorem \eqref{Stokes_theorem} is satisfied. This means we require
\begin{equation}
\begin{split}\label{Stokes}
    \int_\CM \dt C_{d+1} &= \int_{\ci^+} C_{d+1} - \int_{\ci^-} C_{d+1} + \text{(other boundary terms)} \\
\int_{\ci^\pm \cup i^\pm} \dt C_{d} &= \mp \oint_{\ci_\mp^\pm} C_{d}  .
\end{split}
\end{equation}
Note that the $\pm$ signs above are required since $\ci^\pm$ are future/past boundaries of $\CM$ respectively, and $\ci^\pm_\mp$ are past/future boundaries of $\ci^\pm$ respectively. It can now be checked that \eqref{Stokes} implies 
\begin{equation}
\begin{split}
\int_{\ci^\pm} C_{d+1} &= -  \frac{1}{2} \int_{\ci^\pm} \dt u \, \dt^d x \left( \lim_{r\to\pm\infty} |r|^d ( \star C_{d+1} )^r  \right)  \\
\int_{\ci_\mp^\pm} C_{d} &=  \frac{1}{2} \int_{\ci_\mp^\pm} \dt^d x \left( \lim_{u\to\mp\infty}  \lim_{r\to\pm\infty}  |r|^d ( \star C_{d} )^{ur}  \right)  . 
\end{split}
\end{equation}

\subsubsection{Phase Space Manifold}

The phase space is a symplectic manifold $(\G,\BOm)$, where $\G$ is a smooth manifold and $\BOm$ is a closed non-degenerate two-form known as the symplectic form, i.e. it obeys
\begin{align}
\Bdt \BOm = 0, \qquad \Bi_\BX \BOm = 0 \quad\implies\quad \BX = 0 \quad\text{for all $\BX \in T\G$.}
\end{align}
As in \cite{He:2020ifr}, we assume both the first and second cohomology groups vanish, i.e. $\CH^1(\G) = \CH^2(\G) = 0$, so that there exists the one-form $\BTh$ called the symplectic potential such that $\BOm = \Bdt \BTh$. Notice that the symplectic potential is defined only up to an exact one-form.

The phase space for a particular field theory living on a spacetime manifold $\CM$ is specified in terms of the dynamical fields $\varphi^i$ in the field theory. Typically, we impose boundary conditions on the fields $\varphi^i$ to ensure finite energy flux through the boundaries of the spacetime manifold. The configuration space $\mfF$ is then defined to be the space of all field configurations that satisfy the boundary conditions. A vector $\BX \in T\mfF$ is defined in component notation to be
\begin{align}
\BX = \sum_i \int_\CM \e\,\BX^i(\varphi^i,\n_\mu\varphi^i,\ldots) \frac{\d}{\d\varphi^i},
\end{align}
where $\BX^i$ is in general a function of the dynamical fields (as well as any possible background fields), the metric, and spacetime coordinates. Given any function $f$ in terms of the fields, we refer to $\BX(f)$ as the variation of $f$ with respect to $\BX$.

To obtain the symplectic form from a Lagrangian, we utilize the covariant phase space formalism. Given a Lagrangian spacetime $(d+2)$-form $\msL$ involving fields $\varphi^i$, we have
\begin{align}
\BX(\msL) = \sum_i \CE_i \BX(\varphi^i) + \dt {\bs \t} (\BX),
\end{align}
where ${\bs\t}$ is known as the symplectic potential current density, and $\CE_i$ are the equations of motion. The solution space $\mfS$ is defined to be the subspace of $\mfF$ where the fields satisfy $\CE_i=0$, and the tangent space $T\mfS$ consists of vector satisfying the linearized equations of motion $\BX(\CE_i) = 0$. In this paper, we restrict ourselves to the solution space $\mfS$, and field configurations that live in $\mfS$ are known as on-shell configurations.

The symplectic current density ${\bs \o}$ is defined to be ${\bs \o} = \Bdt {\bs \t}$. The pre-symplectic potential and pre-symplectic form can then be obtained by respectively integrating ${\bs\t}$ and ${\bs\o}$ over a future-directed $(d+1)$-dimensional Cauchy slice $\S$, i.e.
\begin{align}
\wt\BTh_\S(\BX) = \int_\S {\bs\t} (\BX), \qquad \wt\BOm_\S(\BX,\BY) = \int_\S {\bs\o}(\BX,\BY).
\end{align}
The pre-symplectic form is a closed two-form on $\mfS$, but it is not necessarily invertible. To remedy this, we determine the kernel of $\wt\BOm$, and we identify $\BX \sim \BX'$ if $\BX-\BX' \in \ker\BOm_\S$. The phase space is then $\G \equiv \mfS/\sim$, and we have $\BTh_\S = \tilde\BTh  |_{\G}$ and $\BOm = {\wt \BOm}|_\G$. Note that by construction, $\BOm$ is both closed and non-degenerate on $\G$. At a practical level, the equivalence relation on $\G$ is imposed by a gauge fixing condition $f[\varphi] = 0$, which maps each equivalence class to a particular representative.

\section{Radial Electric Field as Shadow Transform}\label{app:shadow}

In this appendix, we prove that \eqref{Edef}, which we reproduce here for convenience,
\begin{align}\label{Edef_app}
\begin{split}
E^\pm(x) = \begin{cases}
\pm  \frac{g^2}{2(4\pi)^{\frac{d}{2}} \G \left(\frac{d}{2} \right)} (-\p^2)^{\frac{d}{2}-1} \p^a N_a^{(\pm)}(x) & \text{$d$ even} \\
\pm \frac{g^2 (-1)^{\frac{d-1}{2}} \G(d-1) }{8\pi^{d+1} } \displaystyle\int \dt^dy \frac{\p^a N_a^{(\pm)}(y)}{ [ (x-y)^2  ]^{d-1}} & \text{$d$ odd} ,
\end{cases}
\end{split}
\end{align}
can be unified by utilizing the shadow transform so that
\begin{align}\label{app:Edef_final}
E^\pm = \pm \frac{g^2}{4 c_{1,1}}\p^2 \CN^\pm , \qquad \p_a \CN^\pm = \widetilde{\p_a \Nfield^\pm}. 
\end{align}
We recall from \eqref{shadow_def} that the definition of the shadow transform for vector primaries is
\begin{equation}
\begin{split}
{\wt V}_a(x) \equiv \int \dt^d y \frac{ \CI_{ab}(x-y) }{ [ ( x - y )^2 ]^{d-\D}} V^b(y) , \qquad \CI_{ab}(x) = \d_{ab} - 2 \frac{x_ax_b}{x^2}. 
\end{split}
\end{equation}
This satisfies $\wt{\wt{V_a}}(x) = c_{\D,1} V_a(x)$, where
\begin{equation}
\begin{split}
\label{c11}
c_{\D,1} = \frac{\pi^d(\D-1)(d-\D-1) \G(\frac{d}{2}-\D)\G(\D-\frac{d}{2})}{\G(\D+1)\G(d-\D+1)} . 
\end{split}
\end{equation}
Note that if $\D=1$, then one of the terms in the numerator vanishes. If $d$ is odd, this means $c_{1,1} =0$, although we will see that the final result $E^\pm$ is still finite. If $d$ is even, then $\G(1-\frac{d}{2})$ is divergent for $d \geq 2$, and so the numerator of \eqref{c11} is rendered finite. 

We will now show that \eqref{app:Edef_final} reproduces \eqref{Edef_app}. Let us check the odd-dimensional case first. Because we have $c_{1,1}=0$, we will need to carefully regulate the integral. We do so by evaluating the shadow transform for a field with generic $\D$ and then taking the limit $\D \to 1$. The regulated definition of $E^\pm(x)$ then becomes
\begin{align}
\begin{split}
E^\pm(x) &= \pm \lim_{\D\to1} \frac{g^2}{4c_{\D,1}} \p^a \wt{\p_a \Nfield^\pm}(x) \\
&= \pm \lim_{\D \to 1} \frac{g^2}{4c_{\D,1}} \p^a \int \dt^dy \frac{\CI_{ab}(x-y)}{ [ (x-y)^2  ]^{d-\D}} \p^b \Nfield^\pm(y) \\
&= \pm \frac{g^2 \G(d-1)}{4 \pi^d(d-2) \G ( \frac{d}{2}- 1  )\G ( 1 - \frac{d}{2}  )} \int \dt^dy \, \p^2 \frac{1}{ [(x-y)^2 ]^{d-1}} \Nfield^\pm(y) \\
&= \pm \frac{g^2 (-1)^{\frac{d-1}{2}} \G(d-1)}{8 \pi^{d+1}} \int \dt^dy  \frac{ \p^a \Nfield_a^\pm(y)}{ [(x-y)^2 ]^{d-1}},
\end{split}
\end{align}
where in the second equality we used the definition of the shadow transform for an operator of dimension $\Delta$; in the third equality we used ``integration-by-parts'' style manipulation (IBP) to move the partial derivative onto the other term in the integrand, applied the identity \cite{Kapec:2021eug}
\begin{align}\label{pinvar}
\p^b \bigg\{ \frac{\CI_{ab}(x)}{(x^2)^{d-\D}} \bigg\} = \frac{\D-1}{d-\D} \p_a \bigg\{ \frac{1}{(x^2)^{d-\D}} \bigg\}
\end{align}
for any $\D$ not an integer,\footnote{As we're taking the limit $\D \to 1$, we can assume $\D$ is indeed not an integer and hence apply this equation.} and moved the derivative $\p^a$ on $x$ into the integral and changed it into a $y$ derivative; and in the final equality we used IBP again to move all the derivatives back onto $\Nfield^\pm(y)$ and $\Nfield_a^\pm = \p_a \Nfield^\pm$. This is in agreement with \eqref{Edef_app}.

Next, let us check the even-dimensional case in \eqref{Edef_app}.  It will be useful to first prove the following useful identity: 
\begin{align}\label{useful_id}
\begin{split}
\p^b \bigg\{ \frac{\CI_{ab}(x)}{(x^2)^{d-1}} \bigg\} = \frac{2c_{1,1}}{(4\pi)^{\frac{d}{2}} \G  ( \frac{d}{2}  )} (-\p^2)^{\frac{d}{2}-1 }\p_a \d^d(x).
\end{split}
\end{align}
To prove this, we want to write the right-hand side in a way that can be easily regularized. We claim that we can rewrite this identity as
\begin{align}\label{useful_id2}
\begin{split}
\p^b \bigg\{ \frac{\CI_{ab}(x)}{(x^2)^{d-1}} \bigg\} = -\frac{2c_{1,1}}{(4\pi)^{\frac{d}{2}} \G  ( \frac{d}{2}  )^2} (-\p^2)^{d-1} \p_a \log(x^2). 
\end{split}
\end{align}
To see why, we first note for any positive integer $m$ we have
\begin{align}\label{init}
\begin{split}
-\p^2 \bigg\{ \frac{1}{(x^2)^m} \bigg\} &= \p_a \frac{2m x^a}{(x^2)^{m+1}} = 2m  [ d - 2(m+1)  ] \frac{1}{(x^2)^{m+1}}.
\end{split}
\end{align}
Using this recursively implies that for any positive integer $n$ we have
\begin{align}\label{recursive}
\begin{split}
(-\p^2)^n \bigg\{ \frac{1}{(x^2)^m} \bigg\} &= \frac{4^n\G(m+n)\G ( \frac{d}{2} - m ) }{\G(m)\G ( \frac{d}{2} - m - n ) }  \frac{1}{(x^2)^{m+n}} .
\end{split}
\end{align}
It follows
\begin{align}\label{intermed}
\begin{split}
(-\p^2)^{\frac{d}{2} -1} \log(x^2) &= -2(d-2)(-\p^2)^{\frac{d}{2}-2} \left( \frac{1}{x^2} \right) = -  \frac{2^{d-3} (d-2) \G ( \frac{d}{2}-1)^2}{(x^2)^{\frac{d}{2} - 1}},
\end{split}
\end{align}
where in the second equality we applied \eqref{recursive} with $m=1$ and $n= \frac{d}{2}-2$. Now, if we act on this with $-\p^2$ one more time, it is clear from \eqref{init} with $m = \frac{d}{2}-1$ that the result vanishes. However, this is only true if $x \neq 0$, and so acting on both sides of \eqref{intermed} with $-\p^2$ can give something proportional to the delta function. To determine the coefficient, we first note that we can write the delta function as
\begin{align}\label{delta-rep}
\d^d(x) = \frac{\G (\frac{d}{2}+1)}{\pi^{\frac{d}{2}}} \lim_{\e \to 0} \frac{\e^2}{(x^2+\e^2)^{\frac{d}{2}+1}}. 
\end{align}
To check this is indeed the case, note that for $x\neq 0$, the right-hand side obviously vanishes. Furthermore, integrating it over all of space yields
\begin{align}
\begin{split}
\frac{\G (\frac{d}{2}+1)}{\pi^{\frac{d}{2}}}  \lim_{\e \to 0} \int \dt^dx \frac{\e^2}{(x^2+\e^2)^{\frac{d}{2}}} &= \lim_{\e\to 0} d \int_0^\infty \dt r\, r^{d-1} \frac{\e^2}{(x^2+\e^2)^{\frac{d}{2}+1}} = 1,
\end{split}
\end{align}
which proves \eqref{delta-rep}. 

Similarly regulating the right-hand side of \eqref{intermed}, we obtain
\begin{align}
\begin{split}
\label{log_der}
(-\p^2)^{\frac{d}{2}}\log(x^2) &= - 2^{d-3} (d-2) \G\left( \frac{d}{2}-1 \right)^2 \lim_{\e\to 0} (-\p^2) \bigg\{ \frac{1}{(x^2 + \e^2)^{\frac{d}{2} - 1}} \bigg\} \\
&= - 2^d \G\left( \frac{d}{2} \right)\G\left( \frac{d}{2}+1 \right)\lim_{\e\to 0} \bigg[ \frac{1}{(x^2+\e^2)^{\frac{d}{2}} } - \frac{x^2}{(x^2+\e^2)^{\frac{d}{2} + 1 }} \bigg] \\
&= - 2^{d} \G\left( \frac{d}{2} \right)\G\left( \frac{d}{2}+1 \right) \lim_{\e\to 0} \frac{\e^2}{(x^2+\e^2)^{\frac{d}{2} + 1}} \\
&= - (4\pi)^{\frac{d}{2}} \G\left( \frac{d}{2} \right) \d^d(x) ,
\end{split}
\end{align}
where in the last equality we used \eqref{delta-rep}. Substituting the delta function in \eqref{useful_id} with this expression, we obtain
\begin{align}\label{useful_id2b}
\begin{split}
\p^b \bigg\{ \frac{\CI_{ab}(x)}{(x^2)^{d-1}} \bigg\} = - \frac{2c_{1,1}}{(4\pi)^{d} \G  ( \frac{d}{2}  )^2} (-\p^2)^{d-1 }\p_a \log(x^2) ,
\end{split}
\end{align}
which is exactly \eqref{useful_id2}, as claimed. This means to prove \eqref{useful_id}, it suffices to prove \eqref{useful_id2b}.

Now, note that $c_{1,1}$ is ill-defined since there are both zeros and divergences in the numerator, so we regulate \eqref{useful_id2b} and rewrite it as
\begin{align}\label{useful_id3}
\begin{split}
	\lim_{\D \to 1} \p^b \bigg\{ \frac{\CI_{ab}(x)}{(x^2)^{d-\D}} \bigg\} &= - \lim_{\D\to 1} \frac{2c_{\D,1}}{(4\pi)^{d} \G \left( \frac{d}{2} \right)^2} (-\p^2)^{d-1 }\p_a \bigg\{ \frac{1}{(\D-1)(x^2)^{1-\D}} \bigg\},
\end{split}
\end{align}
where we noted 
\begin{align}
	\lim_{\D\to1}\frac{1}{(\D-1)(x^2)^{1-\D}} = \frac{1}{\D-1} + \log(x^2),
\end{align}
and the leading $\frac{1}{\D-1}$ is eliminated by the derivative $\p_a$. It follows using \eqref{pinvar} that the left-hand side of \eqref{useful_id3} is
\begin{align}\label{lhs}
\begin{split}
	\lim_{\D \to 1} \p^b \bigg\{ \frac{\CI_{ab}(x)}{(x^2)^{d-\D}} \bigg\} &= \lim_{\D\to1} \frac{\D-1}{d-\D} \p_a \left\{ \frac{1}{(x^2)^{d-\D}} \right\},
\end{split}
\end{align}
whereas the right-hand side of \eqref{useful_id3} is
\begin{align}
\begin{split}
	&- \lim_{\D\to 1} \frac{2c_{\D,1}}{(4\pi)^{d} \G \left( \frac{d}{2} \right)^2} (-\p^2)^{d-1 }\p_a \bigg\{ \frac{1}{(\D-1)(x^2)^{1-\D}} \bigg\} \\
	&\qquad =  - \lim_{\D \to 1} \frac{2c_{\D,1}}{(4\pi)^{d} \G \left( \frac{d}{2} \right)^2 (\D-1)} \frac{4^{d-1} \G(d-\D)\G\left( \frac{d}{2} + \D - 1 \right)}{\G(1-\D) \G\left( \D - \frac{d}{2}  \right) } \p_a \left\{ \frac{1}{(x^2)^{d-\D} } \right\} \\
	&\qquad = \lim_{\D\to 1} \frac{\D-1}{d-1} \p_a \left\{ \frac{1}{(x^2)^{d-\D}} \right\} ,
\end{split}
\end{align}
where in the first equality we used \eqref{recursive} with $n = d-1$ and $m = 1-\D$, and in the last equality we substituted in $c_{\D,1}$ from \eqref{c11} and then expanded about $\D=1$. This is exactly \eqref{lhs}, thus proving \eqref{useful_id3}. As \eqref{useful_id3} is a rewriting of \eqref{useful_id2b}, which as we argued above is equivalent to \eqref{useful_id}, this completes the proof of \eqref{useful_id}.

Having proved the identity, we return to evaluating $E^\pm$ for the even-dimensional case, which is given in \eqref{app:Edef_final} to be
\begin{equation}
\begin{split}
E^\pm &= \pm \frac{g^2}{4 c_{1,1}} \p^a \wt{\p_a \Nfield^\pm} = \pm \frac{g^2}{4 c_{1,1}} \p^a \int \dt^d y  \left\{ \frac{ \CI_{ab}(x-y) }{ [ ( x - y )^2 ]^{d-1}} \right\} \p^b \Nfield^\pm (y) .
\end{split}
\end{equation}
It follows
\begin{equation}
\begin{split}
    E^\pm &= \mp \frac{g^2}{4c_{1,1}} \p^a \int \dt^dy \, \p^b \bigg\{ \frac{\CI_{ab}(x-y)}{[(x-y)^2]^{d-1}} \bigg\} \Nfield^\pm(y) \\
    &= \pm  \frac{g^2}{2(4\pi)^{\frac{d}{2}} \G  ( \frac{d}{2}  )^2} \int \dt^d y  (-\p^2)^{d-1} \p^a\p_a \delta^d ( x - y )  \Nfield^\pm (y)  \\
&= \pm \frac{g^2}{2(4\pi)^{\frac{d}{2}} \G  ( \frac{d}{2}  )}  (-\p^2)^{\frac{d}{2}-1} \p^a \Nfield_a^\pm (x) ,
\end{split}
\end{equation}
where in the first equality we used IBP, in the second equality we used \eqref{useful_id} and moved the derivative $\p^a$ on $x$ into the integral and changed it into a $y$ derivative, and in the final equality we used IBP again to move derivatives back onto $\Nfield^\pm(y)$.
This is in agreement with \eqref{Edef_app}, and completes the proof that \eqref{app:Edef_final} is indeed correct for both odd and even-dimensional cases.

\bibliography{YMbib}{}
\bibliographystyle{utphys}

\end{document}